\documentclass[twoside]{article}

\def\beq{\begin{equation}} \def\eeq{\end{equation}} 
\def\three{\,{}^{(3)}\kern-1.5pt} \def\four{\,{}^{(4)}\kern-1.5pt}
\def\BTs{\hbox{BTs}}
\def\Lie{\pounds}

\def\matbf#1{\mbox{\boldmath$#1$}}
\def\bfe{{\matbf{e}}}
\def\bfg{{\matbf{g}}}
\def\bfk{{\matbf{k}}}
\def\bfS{{\mathbf{S}}}
\def\bfK{{\matbf{K}}}
\def\bfbeta{{\matbf\beta}}
\def\bfone{{\matbf1}}

\def\oversymbol#1#2{\vbox{\ialign{##\crcr \hfil$#1$\hfil\crcr
   \noalign{\kern1pt\nointerlineskip}%
   \hbox{$\hfil\displaystyle#2\hfil$}\crcr}}}
\def\overcirc#1{\oversymbol{\scriptstyle\kern.5pt \circ}{#1}}

\begin{document}

\setbox43=\vbox{\small\noindent
from the 2004 Elba Conference in honor of\\
 Yvonne Choquet-Bruhat's 80th birthday:\\
{\it Analysis Manifolds and Geometric Structures in Physics\/}\\
(G. Ferrarese, Ed.), Bibliopolis, Naples, 2005\\
and Nuovo Cim. B 119, 697--715 (2004)}

\title{
The densitized lapse (``Taub function") and the Taub time gauge in cosmology}
\author{Robert T. Jantzen\protect\footnote{robert.jantzen@villanova.edu}\\
Department of Mathematical Sciences\\ 
Villanova University, Villanova, PA 19085-1699
}
\date{}
\maketitle

\begin{abstract}
The role of the Taub time gauge in cosmology is linked to the use of the densitized lapse function instead of the lapse function in the variational principle approach to the Einstein equations. The spatial metric variational equations then become the Ricci evolution equations, which are then supplemented by the Einstein constraints which result from the variation with respect to the densitized lapse and the usual shift vector field.
In those spatially homogeneous cases where the least disconnect occurs between the general theory and the restricted symmetry scenario, the recent adjustment of the conformal approach to solving the initial value problem resulting from densitized lapse considerations is seen to be inherent in the use of symmetry-adapted metric variables. 
The minimal distortion shift vector field is a natural vector potential for the new York thin sandwich initial data approach to the constraints, which in this case corresponds to the diagonal spatial metric gauge. For generic spacetimes, the new approach suggests defining a new minimal distortion shift gauge which agrees with the old gauge in the Taub time gauge, but which also makes its defining differential equation agree with the vector potential equation for solving the supermomentum constraint in any time gauge.
\end{abstract}

\vglue-2\baselineskip\vskip -6in\vbox{\vss\box43\vss}\vskip6in 

\section{Introduction}

In the 1950s, pioneering work by Lichnerowicz \cite{lich}, Choquet-Bruhat \cite{cb55} and Dirac \cite{dirac} led to the Hamiltonian analysis of general relativity by Arnowitt, Deser and Misner (ADM) \cite{adm} using their ADM decomposition of the spacetime metric variables ($\four g_{\alpha\beta}$) into a spatial metric ($g_{ab}$) on a spacelike slicing of spacetime threaded by time coordinate lines whose tangent determines the lapse function ($N$) and shift vector field ($\beta^a$) variables, terminology introduced by Wheeler \cite{beloha}. More recently the densitized lapse $\alpha=N/g^{1/2}$, namely the lapse function divided by the square root of the spatial metric determinant (a spatial density of weight $-1$), has emerged to play a key role in better understanding the gravitational field dynamics and constraints in general relativity, especially in work by Choquet-Bruhat and York and their collaborators during the past decade (see \cite{acby99,cby03} for a complete list of references, including Teitelboim \cite{tei82,tei83}, Ashtekar \cite{ashtekar} and Frittelli \cite{frittelli}). In addition to the name densitized lapse, this function has been called the lapse density, slicing density, slicing function and lapse antidensity (because of the weight $-1$). York has suggested calling it simply the Taub function in honor of the man who put it to its first significant use in general relativity at the very beginning of the 1950s \cite{taub}. We adopt this name here and follow York's new convention of letting $\alpha$ denote it instead of the lapse function $N$ as in his older work.

These new variables of the $3+1$ decomposition of the spacetime metric represent local changes of the metric variables which lead to new linear combinations of the field equations in the Lagrangian or Hamiltonian approach. The original spacetime metric variables $\four g_{\alpha\beta}$ lead to the field equations in (contravariant) Einstein form, equating the Einstein tensor to the gravitational constant times the matter energy-momentum tensor
\beq
  \four G^{\alpha\beta} = \kappa T^{\alpha\beta}\ .
\eeq
Changing to the lapse, shift and spatial metric variables in the variational approach automatically splits these field equations orthogonally into the spatially projected Einstein equations (the ``evolution equations") and the Einstein constraints. The further change of variables to the Taub function, shift and spatial metric variables reshuffles these equations again to yield the evolution equations in Ricci form, namely the spatial projection of the Ricci form of the Einstein equations
\beq
  \four R^{\alpha\beta} 
 = \kappa (T^{\alpha\beta}-\frac12 \four g^{\alpha\beta} T^\gamma{}_\gamma)
\ ,
\eeq
and the same Einstein constraints. 

This mixed form of the Einstein equations (Ricci evolution equations plus Einstein constraints) has long been known to be rather useful for studying various problems in general relativity. It was used by Darmois \cite{darmois} during the period from the 1920s to the 1930s and continued by his student Lichnerowicz starting in the 1940s and by Choquet-Bruhat in the 1950s and beyond, including York in his review article on the kinematics and dynamics of general relativity in 1979 \cite{york79}.
Most recently it has been recognized that 
with the Ricci evolution equations combined together with the additional evolution equations for the constraint functions coming from the Bianchi identities and source field equations one has a symmetric hyperbolic system for the constraints with possibly important consequences for numerical solution since the constraints then propagate in a causal domain of dependence in contrast with the corresponding system using the Einstein evolution equations \cite{frittelli}." 

The Taub function indirectly made its appearance in Taub's classic paper \cite{taub51} introducing Bianchi cosmology in 1951, where he finds his Taub solution of the diagonal vacuum Bianchi type II case in the time gauge $N=g^{1/2}$,  namely that the Taub function is unity: $\alpha  = 1$.
Misner  \cite{misner70} used this same time gauge (see his Eq.~(44), recalling that $e^{-\Omega} = g^{1/6}$) which he called the supertime time gauge, although a slight (affinely related) variation of it $\alpha = 12$ is more appropriate in the full Hamiltonian analysis since it makes the kinetic energy term in the gravitational Hamiltonian for the diagonal Bianchi models correspond exactly equal to the one associated with the 3-dimensional Minkowski metric in terms of the natural logarithmic isotropy/anisotropy variables $\beta^0=-\Omega$ and $\beta^\pm$ which parametrize the diagonal metric coefficients. This generalizes nicely with the DeWitt inner product on the space of inner products (or simply DeWitt metric \cite{dewitt}) to higher-dimensional spatially homogeneous spacetimes \cite{bob86b}.
Looking back decades later, the Misner (and later Ryan \cite{ryanshepley}) Hamiltonian analysis is a bit hard to follow since they always start with the reduced Hamiltonian in the $\Omega$ time gauge for which 
$\alpha = N/g^{1/2} = 12 (p_0)^{-1}=3/(g^{1/2}{\rm Tr}K)$, where $H=|p_0|$ is the (reduced) Hamiltonian, and $p_0$ is the momentum dual to the isotropy variable $\beta^0=-\Omega=\frac16\ln g$, rather than starting with the Lagrangian/Hamiltonian of the general theory and reducing the latter. In the case of the Taub Bianchi type II vacuuum solution, the Taub time gauge leads to the Hamiltonian problem of scattering in a time-dependent 1-dimensional exponential potential in flat 3-dimensional Minkowski spacetime, which is easily decoupled by a Lorentz transformation into three independent 1-dimensional exponential scattering problems, one with vanishing potential corresponding to free motion, the others easily solved in terms of hyperbolic functions \cite{bob86,bob87,bob88}. For Taub's locally rotationally symmetric Bianchi type IX vacuum solution, the free motion degree of freedom is suppressed by the local rotational symmetry.

For spatially homogeneous spacetimes of any dimension, the term Taub time gauge has been well established to describe the lapse condition
$\alpha  = {\rm constant} $, i.e., the Taub function is fixed to be a constant. It seems reasonable to keep the terminology Taub time gauge to refer to a spacetime constant Taub function function and let the ``generalized Taub time gauge" (or some suitable new name yet to be determined) describe the condition that
$\alpha$ be time-independent in a general inhomogeneous spacetime, i.e., the Taub function is independent of the time coordinate in the associated coordinate system. Of course this includes the Taub time gauge as a special case.
Choquet-Bruhat and Ruggeri found that exactly this latter generalized Taub time gauge converts the $3+1$ split Einstein equations into a hyperbolic system \cite{cbr83} when expressed as a second order system in the extrinsic curvature. They noted that when the shift vector field is zero, the generalized Taub time gauge corresponds to a harmonic time coordinate slicing, namely $t_{;\alpha}{}^{;\alpha}=0 $. A short calculation of the spacetime Laplacian of $t$ in $3+1$ variables
\begin{eqnarray}
    t_{;\alpha}{}^{;\alpha} 
&=& N^{-2}[ \partial_t \ln\alpha + N\nabla_a (N^{-1} \beta^a) ]
\nonumber\\
&=& \alpha^{-1} N^{-2}[ \partial_t - \pounds_{\beta}] \alpha
\end{eqnarray}
where the Lie derivative of the weight $-1$ density $\alpha$ \cite{Schouten} is
\begin{equation}
\pounds_{\beta} \alpha = \beta^a \nabla_a \alpha - \alpha \nabla_a \beta^a\ ,
\end{equation}
(which corrects (13.24) of \cite{mfg} and corresponds to Eqs.~(22)-(25) of \cite{acby00} with $f=0$)
shows that the vanishing of $t_{;\alpha}{}^{;\alpha} $ is equivalent to the Taub function being time-independent when the shift vector field $\beta^a$ is zero (or more precisely, when the relative velocity $N^{-1} \beta^a$ between the time coordinate hypersurface normal direction and the 4-velocity tangent to the time coordinate lines is divergence-free in the spatial geometry). This is exactly the case in the Bianchi type II and IX vacuum spacetime solutions of the Einstein equations found by Taub in zero-shift gauge, where the Taub function, already spatially constant as a spatially homogeneous scalar, is a spacetime constant in harmonic time gauge, the gauge he used to solve the equations. 

The Taub time gauge was also used by Belinski, Khalatnikov and Lifshitz \cite{bkl70} (see their Eq.~(3.4)) in their analysis of the change of Kasner epoch due to a single curvature wall, the Bianchi type II approximation to the spatially homogeneous dynamics in the BKL limit approaching the initial singularity in a generic spacetime in which the dynamics becomes pointwise like spatially homogeneous dynamics. It should be noted that they always considered the Ricci form of 
the spatial Einstein equations in their analyses \cite{lk63} 
and they and others in the Russian school at the time
(see \cite{grish72}, for example)
often used the Taub time gauge to state Einstein's equations for Bianchi models in the form of the Ricci evolution equations plus the Einstein constraints. For the diagonal such models they considered, the time derivative terms in the Ricci evolution equations in the Taub time gauge reduce to second time derivatives of the logarithms of the diagonal spatial metric coefficients, the simplest form they can take, corresponding to the flattened DeWitt metric (having scaled away its spatial metric determinant factor) on the space of these logarithmic variables.

The Taub time gauge considerably simplifies the evolution equations by eliminating the extra terms involving the kinetic energy as a multiplier that arise from the differentiation of the factor $g^{-1/2}$ in the kinetic energy part of the Hamiltonian. A calculation equivalent to the explicit vacuum calculation of Anderson and York \cite{andyork98} (see their Eq.~(12)), shows that this is equivalent to obtaining new evolution equations which are only linear combinations of the spatial projection of the Ricci form of the Einstein equations, even in the nonvacuum case. Jantzen \cite{bob86,bob87} (see also \cite{bob86b}) made this claim without showing the calculation, and used the Taub time gauge and Ricci form of the field equations to explain some exact Bianchi cosmological solutions which are easily obtained in the Taub time gauge, for which the evolution equations (see his Eq.~(8) in \cite{bob87}) are particularly simple for a diagonal metric. These two articles both promote the use of the mixed form of the Einstein equations: Ricci form of the evolution equations plus the Einstein constraints, an approach which was exploited in \cite{bob80} for interpreting vacuum solutions of the Ricci evolution equations for which certain multiples of the Einstein constraints are then conserved, leading to nonvacuum solutions of the full equations.

The Taub function proved very useful in investigating the dynamics of the Bianchi cosmological models. The variable $ x  = \alpha^{-1}$,
namely the reciprocal of the Taub function, was introduced already in Jantzen  \cite{bob80}
as a convenient variable which in the Silkos time gauge $N=(g_{33})^{1/2}$ (equivalently $\alpha = (g_{11}g_{22})^{-1/2}$) for certain Bianchi types and initial data has simple exponential or hyperbolic function solutions.
Later the same symbol $x^{-1}$ was explicitly used for the Taub function due to the simplifying properties of the Taub time gauge and Ricci evolution equations in the spatially homogeneous dynamics \cite{bobclaes92,bobclaeskjell93,claesbobkjell95}. Interestingly enough Anderson, Choquet-Bruhat and York also initially used the reciprocal Taub function which they also called $\alpha$ \cite{acby97} (see their Eq.~(66)) before switching to the Taub function itself. Note that $\alpha$ was the symbol originally used by York for the lapse itself.

This same simplifying feature of the Taub time gauge was used by Moncrief \cite{moncrief} in his study of Einstein equations for spacetimes with one Killing vector, although the gauge applies to the Weyl rescaled metric in his fiber bundle fomulation so it is not immediately apparent, as noted explicitly in \cite{bob88} (see section VI). In fact the more general power law lapse conditions, especially in higher-dimensional spacetimes involving dimensional reductions, together with other variable transformations, seem to be the larger setting in which the simplifications of the Taub time gauge extend to other special symmetries.

Teitelboim introduced the Taub function in his study of the Hamiltonian constraints (see his Eqs.~(4.1) and (4.4) in \cite{tei82}, where he uses the symbol $N$ for the Taub function function) to simplify considerably the Feynmann path integral approach to quantization of the gravitational field, since it removes the square root of the spatial metric determinant from the DeWitt metric on the space of spatial metrics in the Hamiltonian formulation of general relativity, simplifying the path integral measure considerably. He calls the constant Taub function time gauge together with zero shift vector field the ``proper time gauge" (see his Eq.~(4.20a) in \cite{tei82}), which is a clear misnomer since this is not associated with a proper time coordinate gauge, acknowledged just before his Eq.~(2.2) in \cite{tei83}, but arose from the analogy with the dynamics of a relativistic point particle. Apparently similar considerations motivated Ashtekar \cite{ashtekar} to introduce the Taub function in his ``new variables" (see his footnote 17).

Section 2 shows how two simple changes of variables, first from the spacetime metric coefficient fields to the lapse, shift and spatial metric and then to the Taub function, shift and spatial metric associated with a given slicing and threading of spacetime, repackage the Einstein equations in terms of constraints and evolution equations when derived from a variational principle.
Section 3 shows how these changes in evolution equations affect the evolution of the constraints.
Section 4 reviews the final version of the conformal approach to the initial value problem of solving the supermomentum constraint using a vector potential, which has been improved by Pfieffer and York \cite{py03}based on Taub function considerations, and suggests extending this improvement to the minimal distortion shift gauge so that its defining differential equation agrees with the vector potential equation.
Section 5 shows how the natural symmetry-adapted decomposition of the metric variables in spatially homogeneous cosmology nicely reflects the new approach.

\section{Local variable changes and the chain rule in the variational approach}

Starting with a timelike foliation of the spacetime and adapted coordinates $\{x^\alpha\} = \{x^0=t,x^a\}$, the spacetime metric line element
\beq
  ds^2 = \four g_{\alpha\beta} dx^\alpha\, dx^\beta
       =-N^2 dt^2 + g_{ab} (dx^a + \beta^a dt)(dx^b + \beta^b dt)
\eeq 
can be re-expressed in terms of the ADM variables: the lapse function $N$, the shift vector field $\beta^a$ and the spatial metric $g_{ab}$ (inverse denoted by $g^{ab}$). This can be thought of as a local change of metric variables
\beq
  (\four g_{00},\four g_{0a}, \four g_{ab})
  =(-(N^2-g_{ab}\beta^a \beta^b), g_{ab}\beta^b, g_{ab})
\rightarrow
  (N, \beta^a, g_{ab})\ ,
\eeq
where the absolute value of the spacetime and spatial metric determinants 
$\four g = |\det(g_{\alpha\beta})|$ and $g=|\det(g_{ab})|$
are related by the square of the lapse  $\four g=N^2 g$.
The associated differential is
\begin{eqnarray}
  &&(d\four g_{00},d \four g_{0a}, d \four g_{ab})\nonumber\\
  &&\quad 
   =(-2 N dN + 2\beta_a d\beta^a +\beta^a\beta^b dg_{ab}, 
   g_{ab}d\beta^b +\delta^{(c}{}_a \delta^{d)}{}_b \beta^b dg_{cd}, 
   dg_{ab})\ ,
\end{eqnarray}
while the contravariant metric variables are
\beq
  (\four g^{00},\four g^{0a}, \four g^{ab})
  =(-N^{-2}, N^{-2}\beta^a, g^{ab}-N^{-2}\beta^a \beta^b)
\ .
\eeq

With the standard variational formulas
\begin{eqnarray}
 \delta \four g^{1/2} 
   &=& \frac12 \four g^{1/2} \four g^{\alpha\beta} \delta \four g_{\alpha\beta}\ ,
\nonumber\\
 \delta \four g^{\alpha\beta} 
   &=& - \four  g^{\alpha\gamma} \four g^{\beta\delta} \delta \four g_{\gamma\delta}\ ,
\nonumber\\
 \delta (\four g^{1/2} \four g^{\alpha\beta} \four R_{\alpha\beta})
   &=&  \four g^{1/2} (\frac12 \four g^{\alpha\gamma} \four g^{\beta\delta} 
                 \four R_{\gamma\delta} 
                     - \four R^{\alpha\beta}) \delta \four g_{\alpha\beta}
\nonumber\\ &&
  + \four g^{1/2} \four g^{\alpha\beta}\delta \four R_{\alpha\beta}\ ,
\end{eqnarray}
the usual scalar curvature gravitational Lagrangian produces the lefthand side of the field equations in Einstein tensor form upon variation
\beq
  L = \four g^{1/2} \four R\ ,\
  \delta L = -\four g^{1/2} \four G^{\alpha\beta} dg_{\alpha\beta} +\BTs
\ ,
\eeq
where ``BTs" (boundary terms) stands for additional divergence terms which integrate out to the boundary of a suitable tubular integration region between two time coordinate slices compatible with the slicing and its adapted coordinates, where variations are fixed and so make no contribution to the field equations under suitable conditions on the metric variables. Of course these boundary terms are themselves important for various reasons, but not for the present discussion and so their details will be omitted. York pointed them out in the $3+1$ approach in 1972 \cite{york72}, while soon after they were analyzed from a different point of view by Regge and Teitelboim \cite{reggetei74}. Wald has discussed them in appendix E of his textbook \cite{wald}. One must add terms to the various variational principles to cancel out unwanted boundary terms on the bounding hypersurface of the spacetime region of variation, since only the metric variables themselves can be held fixed there and not their off-hypersurface derivatives.

This Hilbert Lagrangian in turn differs from the ADM gravitational Lagrangian only by such boundary terms $L_{ADM} = L + \BTs$ and so produces the same left hand side of the field equations
\beq
    \delta L_{ADM} 
    = -N g^{1/2} [ \four G^{ab} \delta\four g_{ab} 
                 + 2\four G^{0b} \delta\four g_{0b}
                 + \four G^{00} \delta\four g_{00} ]
      +\BTs \ .
\eeq
Explicitly,
\begin{eqnarray}
  L_{ADM} &=& N g^{1/2} [ K^a{}_c K^c{}_b -K^a{}_a K^b{}_b + R]
\nonumber\\
      &=& \pi^{ab} \dot g_{ab} - N\mathcal{H} - \beta^a \mathcal{H}_a 
+\BTs\ ,
\end{eqnarray}
where $\dot f = \partial f/\partial t$ is the ordinary time derivative acting on a function (equivalent to $\Lie_{\partial_t}$ when acting on tensors since the coordinates are dragged along), and the extrinsic curvature, canonically conjugate momentum, supermomentum and super-Hamiltonian are, letting a vertical bar denote the spatial covariant derivative 
$X^a{}_{|b} = \nabla_b X^a$),
\begin{eqnarray}\label{eq:KpiH}
K_{ab} &=& -\frac1{2N}[\dot g_{ab} -\Lie_\beta g_{ab}]\ ,\ 
\pi^{ab} = \partial L_{ADM}/\partial \dot g_{ab} 
         = -g^{1/2}[K^{ab} -g^{ab}K^c{}_c]\ ,
\nonumber\\
\mathcal{H}_a &=& 2 N g^{1/2} \four G^0{}_a 
               =  2 g^{1/2} \four G^\bot{}_a
            = -2 g^{1/2}(K^b{}_a - K^c{}_c \delta^b{}_a)_{|b}
            = -2 \pi^b{}_{a|b}\ ,
\nonumber\\
\mathcal{H} &=& -2 N^2 g^{1/2} \four G^{00}
            =  2 g^{1/2} \four G^\bot{}_\bot
\nonumber\\
            &=& g^{1/2} [ K^a{}_c K^c{}_b -K^a{}_a K^b{}_b - R]
            = g^{-1/2} [ \pi^a{}_c \pi^c{}_b -\frac12\pi^a{}_a \pi^b{}_b
                             - R]\ .
\end{eqnarray}
Here the ``perp" indices refer to components along the timelike unit normal vector field to the spacelike slicing
$e_\bot = N^{-1}(\partial_t -\beta^a\partial_a)$  and its sign-reversed index-lowered 1-form $\theta^\bot = N dt$;
note that $K_{ab} = -\frac12 \Lie_{e_\bot} g_{ab} $.
 The first two terms in the super-Hamiltonian are the kinetic energy terms corresponding to the DeWitt metric \cite{dewitt}. 

Simply re-expressing the variation of the ADM Lagrangian using the chain rule for the change to ADM metric variables automatically leads to new linear combinations of the field equations which reflect the spacetime splitting of the symmetric second rank tensor into a space-space projection, a time-space projection and a time-time projection under the orthogonal decomposition of the tangent space into a spatial subspace tangent to the slicing and its timelike normal direction. Contravariant 0 indexed components correspond to the time projection (since $\theta^\bot = -e_\bot^\flat$ is the sign-reversed index-lowered normal), while covariant spatial indices correspond to the spatial projection (since $e_a$ are tangent to the spatial hypersurface of constant time) 
\begin{eqnarray}
    \delta L_{ADM} 
    &=& -N g^{1/2} [ g^{ac} g^{bd} \four G_{cd} \delta g_{ab} 
                 + 2\four G^0{}_b \delta\beta^b
                 -2N \four G^{00} \delta N
      +\BTs] 
\nonumber\\
    &=& [-N g^{1/2} g^{ac} g^{bd} \four G_{cd}] \delta g_{ab} 
                 + \mathcal{H}_b \delta\beta^b
                 +\mathcal{H} \delta N
      +\BTs \ .
\end{eqnarray}
The spatial projection of the Einstein tensor leads to the left hand side of the evolution equations in Einstein (tensor) form, while the remaining field equation left hand sides are the gravitational super-momentum and super-Hamiltonian constraints, namely the Einstein gravitational constraint functions. (Obviously the time-space projection of either the Einstein tensor or Ricci tensor form of the field equations produces the same supermomentum constraint.)

A final change of variables expressing the lapse function in terms of the Taub function $\alpha = N/g^{1/2}$, with differential
\beq
d N = d(\alpha g^{1/2}) = g^{1/2} d\alpha + \alpha g^{1/2} g^{ab} dg_{ab}
\ ,
\eeq
leads to a final shuffling of the field equations
\begin{eqnarray}
    \delta L_{ADM} 
    &=& -N g^{1/2}[g^{ac} g^{bd} \four G_{cd} + g^{ab}\four G^\bot{}_\bot]
                \delta g_{ab}
\nonumber\\ 
        &&+ \mathcal{H}_b \delta\beta^b
        + g^{1/2}\mathcal{H} \delta\alpha
      +\BTs \ .
\end{eqnarray}
The new left hand sides of the gravitational evolution equations are now in a trace-modified Ricci tensor form, as a side calculation shows.

The orthogonal splitting of the trace of the Ricci tensor is
\beq
  \four R = -N^2 \four R^{00} + g^{cd} \four R_{cd}
          = \four R^\bot{}_\bot + g^{cd} \four R_{cd}\ ,
\eeq
which implies that
\begin{eqnarray}
 \four G^\bot{}_\bot 
       &=& -N^2 \four G^{00}
       = (-N^2)(\four R^{00}-\frac12 \four g^{00} \four R)
\nonumber\\
       &=& \frac12( \four R^\bot{}_\bot - g^{cd} \four R_{cd}) 
\end{eqnarray}
and so
\beq
 \four R^\bot{}_\bot = 2\four G^\bot{}_\bot + g^{cd} \four R_{cd}\ .
\eeq
Then the spatial projection of the Einstein tensor satisfies
\begin{eqnarray}
  \four G_{ab} +g_{ab} \four G^\bot{}_\bot
 &=& \four R_{ab}-\frac12 g_{ab} (\four R^\bot{}_\bot + g^{cd} \four R_{cd}
            -2 \four G^\bot{}_\bot)
\nonumber\\
 &=& \four R_{ab}- g_{ab} g^{cd} \four R_{cd}
\nonumber\\
 &=& (\delta^c{}_a \delta^d{}_b - g_{ab} g^{cd}) \four R_{cd} =S_{ab}
\ .
\end{eqnarray}
This transformation associated with index-shifting by the DeWitt metric on the space of symmetric tensors only changes the trace of the spatial projection of the Ricci tensor by the factor $1-3=-2$ and is easily inverted by
\beq
\four R_{ab} 
 = (\delta^c{}_a \delta^d{}_b -\frac12 g_{ab} g^{cd}) S_{cd}\ ,
\eeq
leading to the final formula
\begin{eqnarray}
    \delta L_{ADM} 
    &=& -N g^{1/2}[g^{ac} g^{bd} - g^{ab} g^{cd}]\four R_{cd}
                \delta g_{ab}
\nonumber\\ 
        &&+ \mathcal{H}_b \delta\beta^b
        + g^{1/2}\mathcal{H} \delta\alpha
      +\BTs \ .
\end{eqnarray}
Thus with the new independent variables, the evolution equations are equivalent to the Ricci tensor form of those evolution equations, as shown by Frittelli \cite{frittelli} and by Anderson and York \cite{ay98}.

In spatially homogeneous cosmology, the Taub time gauge makes $\alpha$ constant, so imposing this time gauge on the Lagrangian automatically yields the Ricci evolution equations as the variational equations for the spatial metric. The simplification in the variation which produces these equations is obvious because the $g^{1/2}$ factor in the kinetic energy is removed, and no longer generates additional terms proportional to the kinetic energy (and hence to the super-Hamiltonian constraint function when combined with the corresponding term arising from the potential energy term, namely the spatial scalar curvature term).

\section{Evolution of the Constraints}

Working with the original coordinate frame shows clearly the simple changes of variable and their subsequent effect on the form of the variational equations. However, the orthogonal decomposition of tensors is very cumbersome since the frame is not adapted to the spacetime splitting parallel and perpendicular to the time coordinate hypersurfaces. By replacing the time coordinate derivative in the coordinate frame by its normal component, one obtains a partially orthogonal frame adapted to this splitting which makes it trivial to split tensor equations
\begin{eqnarray}
  e_0 = \partial_t-N^a \partial_a= N e_\bot\ ,\ e_a = \partial_a\ ,
  \theta^0 =  dt = N^{-1} \theta^\bot\ ,\ \theta^a = dx^a +N^a dt\ .
\end{eqnarray}
This is called the Cauchy adapted frame for the foliation by Choquet-Bruhat \cite{cby03}. Normalizing the normal and its dual one-form is even more convenient for calculation.

One can generalize these frames to adapted computational frames \cite{york79} by letting $e_a$ be any spatial frame (each frame vector tangent to the time slices) which is Lie dragged along the time coordinate lines
\beq
  [e_a,e_b] = C^c{}_{ab} e_c\ ,\ 
  [\partial_t, e_a] = 0\ .
\eeq
This is necessary in order to apply the present discussion to symmetry adapted frames in the case of spatially homogeneous spacetimes, for example.

The behavior of the constraints depends on which evolution equations one uses. One splits the twice contracted Bianchi identities (divergence of Einstein equals zero) by evaluating components in one of the above adapted frames. 
Let $M^{\alpha\beta} = \four G^{\alpha\beta}-\kappa T^{\alpha\beta}$. Then
\begin{eqnarray}
M^\alpha{}_{\bot ; \alpha} &=&  
g^{-1/2} \partial_\bot (g^{1/2} M^\bot{}_\bot ) 
  + N^{-1}(N M^c{}_\bot)_{|c}
  +M^c{}_d K^d{}_c 
\ ,\nonumber\\
M^\alpha{}_{   a ; \alpha} &=&  g^{-1/2} \Lie_{e_\bot}( g^{1/2} M^\bot{}_a )
  -N^{-1}N_{|a} M^\bot{}_\bot
  + N^{-1}(N M^c{}_a )_{|c}
\ .
\end{eqnarray}
Provided that the matter equations of motion are satisfied, the divergence of the energy-momentum tensor is zero and hence these expressions must vanish. In terms of the total super-Hamiltonian and supermomentum
\beq
  \mathcal{H}^{\rm( total)} = 2 g^{1/2} M^\bot{}_\bot\ ,\
  \mathcal{H}^{\rm( total)}_a = 2 g^{1/2} M^\bot{}_a\ ,
\eeq
these become
\begin{eqnarray}
g^{-1/2} \partial_\bot (\mathcal{H}^{\rm( total)} ) 
 &=& -  N^{-1}(N g^{-1/2} \mathcal{H}^{\rm( total)}{}^c)_{|c}
  - 2 M^c{}_d K^d{}_c 
\ ,\nonumber\\
  g^{-1/2} \Lie_{e_\bot}( \mathcal{H}^{\rm( total)}_a )
 &=& N^{-1}g^{-1/2} N_{|a} \mathcal{H}^{\rm( total)}
  - 2 N^{-1}(N M^c{}_a )_{|c}
\ .
\end{eqnarray}
When the Einstein evolution equations $M_{ab}=0$ hold, 
explicitly
\begin{eqnarray}
 N g^{1/2} M^{ab} &=& \Lie_{e_0} \pi^{ab}
 + 2N g^{-1/2}(\pi^a{}_c \pi^{cb}-\frac12 \pi^{ab}\pi^c{}_c)
\nonumber\\
&&  -\frac12 N g^{-1/2} g^{ab}(\pi^d{}_c \pi^c{}_d -\pi^c{}_c \pi^d{}_d)
 + N g^{1/2} (G^{ab}-\kappa T^{ab})
\nonumber\\
 &&+ N g^{1/2} (N^{|ab}-g^{ab}N^c{}_{|c}) 
 = 0\ ,
\end{eqnarray}
the constraint evolution equations become very simple and confirm that if the constraint functions are zero on the initial hypersurface, they continue to remain so during the evolution.

To see the consequences for the constraints when instead imposing the Ricci evolution equations
\begin{eqnarray}
   \four R_{ab} &=& \kappa(T_{ab} -\frac12 g_{ab} T)    
\leftrightarrow
  \four R_{ab} - g_{ab}\four R^c{}_c = \kappa(T_{ab} + g_{ab} T^\bot{}_\bot)
\nonumber\\
 \leftrightarrow
0&=&M_{ab} +g_{ab}  M^\bot{}_\bot
\nonumber\\
&=& \Lie_{e_0} \pi^{ab}
 + 2N g^{-1/2}(\pi^a{}_c \pi^{cb}-\frac12 \pi^{ab}\pi^c{}_c)
\nonumber\\
&&
 + N g^{1/2} (R^{ab}-g^{ab}R^c{}_c -\kappa T^{ab} -\kappa g^{ab} T^\bot{}_\bot)
\nonumber\\
 &&+ N g^{1/2} (N^{|ab}-g^{ab}N^c{}_{|c}) 
\ ,
\end{eqnarray}
one must re-express the above equations in terms of the combination
$M_{ab} +g_{ab}  M^\bot{}_\bot$. Using the identity
\beq
  K = K^c{}_c
    =-\frac12 g^{ab}\Lie_{e_\bot} g_{ab} =  -\partial_\bot g^{1/2}\ , 
\eeq
one finds
\begin{eqnarray}
M^\alpha{}_{\bot ; \alpha} &=&  
g^{-1/2} \partial_\bot (g^{1/2} M^\bot{}_\bot ) 
  + N^{-1}(N M^c{}_\bot)_{|c}
  + M^\bot{}_\bot g^{-1/2} \partial_\bot (g^{1/2}) 
\nonumber\\
&&  + [(M^c{}_d +\delta^c{}_d M^\bot{}_\bot) K^d{}_c] 
\ ,\nonumber\\
M^\alpha{}_{   a ; \alpha} &=&  g^{-1/2} \Lie_{e_\bot} ( g^{1/2} M^\bot{}_a )
  -N^{-1}N_{|a} M^\bot{}_\bot
  + N^{-1}(-N M^\bot{}_\bot)_{|a} 
\nonumber\\
&& + [ N^{-1}(N M^c{}_a +N \delta^c{}_a M^\bot{}_\bot)_{|c}]
\ ,
\end{eqnarray}
or
\begin{eqnarray}
M^\alpha{}_{\bot ; \alpha} &=&  
g^{-1} \partial_\bot (g M^\bot{}_\bot ) 
  + N^{-1}(N M^c{}_\bot)_{|c}
\nonumber\\
&&  + [(M^c{}_d +\delta^c{}_d M^\bot{}_\bot) K^d{}_c] 
\ ,\nonumber\\
M^\alpha{}_{   a ; \alpha} &=&  g^{-1/2} \Lie_{e_\bot} ( g^{1/2} M^\bot{}_a )
  + N^{-2}(-N^2 M^\bot{}_\bot)_{|a} 
\nonumber\\
&& + [ N^{-1}(N M^c{}_a +N \delta^c{}_a M^\bot{}_\bot)_{|c}]
\ .
\end{eqnarray}

Again provided that the matter equations of motion are satisfied,
one then has the result of Anderson and York \cite{ay98} trivially extended to the nonvacuum case
\begin{eqnarray}  
g^{-1} \partial_\bot (g^{1/2} \mathcal{H}^{\rm( total)} ) 
 &=&  - N^{-1}(\alpha \mathcal{H}^{\rm( total)}{}^c)_{|c}
\nonumber\\
&&  - 2[(M^c{}_d +\delta^c{}_d M^\bot{}_\bot) K^d{}_c] 
\ ,\nonumber\\
  g^{-1/2} \Lie_{e_\bot} ( \mathcal{H}^{\rm( total)}_a)
 &=&  N^{-2}(\alpha^2 g^{1/2} \mathcal{H}^{\rm( total)})_{|a} 
\nonumber\\
&& - 2[ N^{-1}(N M^c{}_a +N \delta^c{}_a  M^\bot{}_\bot)_{|c}]
\ .
\end{eqnarray}
Now when the Ricci evolution equations hold, the square bracketed terms are zero. However, as noted in the introduction, the combined Ricci evolution plus constraint equations supplemented by the propagation equations for the constraints themselves form a symmetric hyperbolic system, with the constraints propagating in a stable manner that is important for numerical solution, in contrast with the corresponding Einstein evolution system  \cite{frittelli}.

While the vacuum Einstein evolution equations are changed by the introduction of an energy-momentum tensor whose spatial part is nonzero, the Ricci evolution equations can remain unchanged if the spatial energy-momentum tensor is related to the energy-density by
\beq
T_{ab} +g_{ab}  T^\bot{}_\bot = 0\ .
\eeq
For a perfect fluid source, the energy-momentum tensor is
\beq
  T_{\alpha\beta} = \rho u_\alpha u_\beta + p (u_\alpha u_\beta + \four g_{\alpha\beta})\ .
\eeq
If this fluid is not tilted, i.e., is flowing orthogonally to the time hypersurfaces (4-velocity $u_\alpha = n_\alpha\rightarrow u^a=0,u^\bot=1$),
then $T_{ab} = p g_{ab}$, $T^\bot{}_\bot = -\rho$, so if the fluid is stiff ($p=\rho$), then this condition is satisfied and the Ricci evolution equations are not changed. 

Thus if one has a solution of the vacuum Ricci evolution equations and the vacuum supermomentum constraint, one can use the super-Hamiltonian constraint function to define the energy-density $\rho$ of a nontilted stiff perfect fluid when the gravitational super-Hamiltonian is negative, and it will satisfy $\partial_\bot (\rho g) = 0$, leading to a conserved quantity for the equivalent nonvacuum system. 

This is the stuffing operation for spatially homogeneous and spatially self-similar spacetimes \cite{bob80}.
For such spacetimes one can also extend this re-inter\-pretation of the vacuum Ricci evolution equation solutions with nonzero constraint functions to the tilted case where the tilt is along an invariant direction, say $e_3$. Decoupling and then solving the evolution equations requires that $\alpha$ be independent of $g_{33}$, a convenient choice being 
$\alpha = (\det g^{(2)})^{1/2}$, where $g^{(2)}$ is the determinant of the remaining block of the spatial metric. One can then interpret solutions of the Ricci evolution equations with nonzero $\mathcal{H}$ and $\mathcal{H}_3$ as a tilted stiff perfect fluid. In the variational point of view, this time gauge leads to another set of evolution equations that are better suited to the particular symmetries of this class of models. 

More generally the power law lapse time gauges \cite{bob88} extend this kind of simplication to other scenarios in spatially homogeneous (and spatially self-similar) cosmology. In fact, almost all exact solutions of this type can be found by choosing the lapse function to be a power law function of appropriate metric components in order to obtain new evolution equations through the variational (or extended variational) approach which decouple for the appropriate choice of metric variables and thus admit solution \cite{claesbobkjell95}.

\section{The New Approach to the Initial Value Problem}

To discuss the conformal approach to the initial value problem, we follow the notation of Pfeiffer and York \cite{py03}. Let an overbar denote the physical metric variables, which are related by a conformal rescaling to the unphysical variables without an overbar. The Taub function and the shift vector field and the trace $\tau$ of the extrinsic curvature are not transformed
\begin{eqnarray}
\bar\alpha &=& \alpha\ ,\
\bar \beta^a = \beta^a\ ,\
\bar g_{ab} = \phi^4 g_{ab}\ ,\   
\bar g{}^{1/2} = \phi^6 g{}^{1/2}\ ,\  
\bar N = \phi^{6} N\ ,
\nonumber\\
  \bar K{}^a{}_b 
   &=& -\frac1{2N} (\bar g{}^{ac}\partial\bar g_{cb}/\partial t 
                 -\bar g{}^{ac}\Lie_\beta \bar g_{cb} )
   = -\frac1{2\alpha \bar g{}^{1/2}} (\bar g{}^{ac}\partial\bar g_{cb}/\partial t 
                 -\bar g{}^{ac}\Lie_\beta \bar g_{cb} )
\nonumber\\
   &=& \bar A{}^a{}_b +\frac13 \tau \delta^a{}_b
\ ,
\end{eqnarray}
where $\bar A{}^a{}_b = \bar K{}^{(TF)}{}^a{}_b$ is the tracefree part of the extrinsic curvature.
The fixing of the Taub function under the conformal transformation is motivated by the nice properties of the Einstein equations that occur when this quantity is fixed.

The mixed form of the Lie derivative term appearing in the extrinsic curvature has the transformation law
\beq
  \bar g{}^{ac} \Lie_\beta \bar g_{cb}
     =  g{}^{ac} \Lie_\beta  g_{cb} + 4 \ln\phi_{,c} \beta^c \delta^a{}_b\ ,
\eeq
so its tracefree part is invariant
\beq
  [\bar L \beta]^a{}_b =  [\bar g{}^{ac} \Lie_\beta \bar g_{cb}]^{(TF)}
     =  [g{}^{ac} \Lie_\beta  g_{cb}]^{(TF)}
  = [L \beta]^a{}_b 
\ .
\eeq
The same is true of the tracefree part of the time derivative term
\beq
 [\bar g{}^{ac} \partial\bar g_{cb}/\partial t ]^{\rm(TF)}
     =  [g{}^{ac} \partial g_{cb}/\partial t]^{\rm(TF)}\ ,
\eeq
suggesting that the conformal transformation of the tracefree part of the extrinsic curvature should be due entirely to the factor of the lapse. This is in fact reinforced by the conformal transformation properties of the divergence operator appearing in the supermomentum constraint. This constraint
\beq
\bar\nabla_b (\bar K{}^b{}_a - \bar K{}^c{}_c \delta^b{}_a) = \bar j_a\ ,
\eeq
can be thought of as a condition on the divergence of the tracefree part of the extrinsic curvature, determining it in terms of the trace $\tau$ and the source current 
$\bar j_a= \bar T{}^\bot{}_a$
\beq\label{eq:smcon}
\bar\nabla_b \bar A{}^b{}_a = \frac23 \nabla_a \tau + \bar j_a\ .
\eeq

To solve the supermomentum constraint in the Hamiltonian picture, namely in terms of the initial variables $\bar g_{ab}$, $\bar K_{ab}$ and $\bar j_a$, one can conveniently choose a conformal representation of the tracefree part of the extrinsic curvature. Letting $\bar\nabla_a$ and $\bar\nabla_a$ be the spatial covariant derivatives associated with the two spatial metrics, the transformation of the divergence of a symmetric tensor $S^{ab}$ is easily evaluated
\begin{eqnarray}
\bar S{}^{ab} &=& \phi^{x-4} S^{ab}  \quad {or}\quad
\bar S{}^a{}_b = \phi^{x} S^a{}_b \nonumber\\
&\rightarrow&  
\bar\nabla_b \bar S{}^b{}_a =\phi^x (\nabla_b S^b{}_a +(x+6)S^b{}_a\nabla_b \ln\phi
                           -2 S^b{}_b \nabla_a \ln\phi) \ .
\end{eqnarray}
Thus picking $x=-6$ makes the divergence of a tracefree such tensor also transform by a conformal factor, namely by the weight $-6$ for the divergence 1-form. This corresponds exactly to the transformation due to the reciprocal lapse factor above. One can decompose the tracefree part of any symmetric tensor into a transverse traceless part (zero divergence) and a pure divergence part using the tracefree Lie derivative operator.
However, the covariant form of the divergence of the tracefree Lie derivative operator is invariant (if the vector doing the derivative is invariant), so one must include an additional transforming factor in it to get the two pieces to transform consistently
\begin{eqnarray}
  \bar A{}^a{}_b &=& \bar A_{(TT)}{}^a{}_b + \bar\sigma^{-1} [\bar L Y]^a{}_b\ ,\nonumber\\
  \bar A_{(TT)}{}^a{}_b &=& \phi^{-6} A_{(TT)}{}^a{}_b\ ,\
  [\bar L Y]^a{}_b =  L Y^a{}_b\ ,\
  \bar\sigma= \phi^{6} \sigma\ , 
\end{eqnarray}
so that the vector potential equation takes the form
\beq\label{eq:VP1}
\bar\nabla_b (\bar\sigma^{-1} [\bar L Y]^b{}_a ) 
= \bar\nabla_b \bar A{}^b{}_a
\eeq
in terms of the barred variables, where the right hand side can be replaced using the supermomentum constraint Eq.~(\ref{eq:smcon}).
This is the final improved conformal approach discovered by York \cite{py03}.
The only obvious candidate for an extra initial data variable that transforms as $\sigma$ should is the metric volume factor $g{}^{1/2}$ or the lapse $N$, which has the same transformation properties when the Taub function is fixed. 
A convenient choice is $\sigma =2N$. This removes the previous ambiguity in the decomposition process (decomposing before or after the conformal transformation) and is much more satisfying. 

This additional factor of $\sigma$ inside the divergence operator corresponds to orthogonality of the splitting of the tracefree extrinsic curvature into transverse traceless and longitudinal parts under an inner product which includes the lapse as a multiplier of the spatial volume integration factor $g^{1/2}$, i.e., in the combination $\four g^{1/2}=N g^{1/2}$ (see the discussion after Eq.~(4.6) of Pfeiffer and York \cite{py03}). This is easily seen by an integration by parts in which the extra factor of the lapse multiplying $g^{1/2}$ cancels the factor of the lapse multiplying the conformal Lie derivative before that integration by parts, allowing the covariant derivative to flip over to the transverse traceless part yielding zero.

The initial value problem takes place at one initial hypersurface, but if one has an evolving spacetime solution of the Einstein equations, one can see what this decomposition looks like in terms of the evolving spatial metric, which in the limit near the initial hypersurface takes us to the ``thin sandwich data" approach \cite{york98} to the initial value problem using the spatial metric and its time derivative $(\bar g_{ab},\dot {\bar g}_{ab})$ as variables instead of the spatial metric and the extrinsic curvature $(\bar g_{ab},\bar K_{ab})$, with
\beq
  \dot {\bar g}_{ab} = -2 \bar N \bar K_{ab} + \Lie_\beta \bar g_{ab} \ ,\
  \bar K_{ab} = -\frac1{2\bar N}(\dot {\bar g}_{ab} -\Lie_\beta \bar g_{ab})\ ,
\eeq
giving the transformation of variables in both directions. One can then unambiguously translate the decomposition of the tracefree part of the extrinsic curvature into a decomposition of the tracefree part of the metric velocity, or the ``conformal metric velocity"
\beq
  \dot {\bar g}_{ab}{}^{\rm(TF)} 
  = -2\bar N \bar A_{ab} + [\bar L\beta]_{ab} \ ,\
  \bar A_{ab} 
  = -\frac1{2\bar N}(\dot {\bar g}_{ab}{}^{\rm(TF)} -[\bar L\beta]_{ab})\ .
\eeq
The proper time velocity is also useful
\beq
 \overcirc {\bar g}_{ab}{}^{\rm(TF)}
  = \bar N{}^{-1} \dot {\bar g}_{ab}{}^{\rm(TF)} 
  = -2 \bar A_{ab} + \bar N{}^{-1} [\bar L\beta]_{ab} \ .
\eeq 

The obvious next question to ask would be, how do we fix the two parts (transverse and longitudinal) of the tracefree part of the extrinsic curvature in either point of view? In the thin sandwich picture how is the potential vector field $Y^a$ related to the shift vector field $\beta^a$? 

Given $\bar \sigma$ and $\bar g_{ab}$, starting from any tracefree symmetric tensor $\bar C_{ab}$, one can remove its divergence to get a transverse traceless symmetric tensor 
\beq
   \bar A_{\rm(TT)}^{ab} = \bar C^{ab}-\bar\sigma^{-1} [\bar L V]^{ab}\ ,\
  \bar\nabla_b (\bar\sigma^{-1} [\bar L V]^{ab}) = \bar\nabla_b \bar C^{ab}
\eeq
which then determines the transverse traceless part of the barred extrinsic curvature by the conformal rescaling. The longitudinal part $[\bar L Y]^{ab}$ of the tracefree extrinsic curvature is then determined by the supermomentum constraint (\ref{eq:smcon}), from which the transverse term drops out, leading to 
\beq\label{eq:VP2}
        \bar\nabla_b [ \bar\sigma^{-1} \bar L Y]^b{}_a 
  = \bar\nabla_b \bar A{}^b{}_a
  = \frac23 \nabla_a \tau + \bar j_a\ .
\eeq
The subtracted divergence part can then be combined with the vector potential term
\beq
  \bar A{}^{ab} = \bar C{}^{ab}+\bar\sigma^{-1} [\bar L (Y-V)]^{ab}\ ,\
\eeq

In the thin sandwich picture, we can make the identifications 
\beq
\bar C_{ab} = -\frac1{2\bar N} [\dot {\bar g}_{ab}]^{\rm(TF)}\ ,\
\bar\sigma^{-1} [\bar L (Y-V)]_{ab} = -\frac1{2\bar N} [\bar L\beta]_{ab}
\ .
\eeq
Thus the shift is related to the original vector potential by
\beq
   \beta^a = Y^a-V^a\ .
\eeq
Suppose we take a spacetime sliced and threaded in zero shift gauge $\beta^a=0$. This implies that the vector potential $Y^a$ must equal the vector $V^a$ which generates the divergence of the tracefree part of the conformal metric velocity. On the other hand one can choose the latter vector to be zero, insisting that the conformal metric velocity be transverse, which forces the shift to equal the vector potential. Both possibilities describe the two most useful spatial gauge choices for the Bianchi type IX spacetimes where the spatial diffeomorphism group is exactly the symmetry group $SO(3,R)$ of the rigid body problem used as an analogy for understanding the spatial diffeomorphism gauge freedom of generic spacetimes by Fischer and Mardsen \cite{fismar}. Zero-shift gauge corresponds to space-fixed coordinates in the rigid body problem, while the transverse gauge corresponds to body-fixed coordinates, with the spatial metric diagonalized exactly as for the moment of inertia tensor in the rigid body problem.

One can also compare the vector potential equation (see Eqs.~(\ref{eq:VP1}), (\ref{eq:VP2})), which is equivalent to 
\beq
    \bar\nabla_b (-\bar A{}^b{}_a + \bar\sigma^{-1} \bar L Y)=0\ ,
\eeq
with the minimal distortion equation of Smarr and York \cite{smarryork}
\beq
 \bar\nabla_b (-\bar\sigma\bar A{}^b{}_a +  \bar L \beta)=0\ .
\eeq
These differ only by the overall multiplicative factor $\bar\sigma$ inside the divergence operation. If $\bar \sigma =2\bar N$ is spatially constant or if the Taub time gauge (constant $\alpha$) is used so that the spatially covariant constant factor of $\bar N =\alpha\bar g{}^{1/2}$ can pass outside the divergence operation, then these two equations coincide and the vector potential for the conformal metric velocity is also a minimal distortion shift vector field, which can then be used to change the spatial gauge to the minimal distortion gauge in order to minimize the time-rate of change of the magnitude of the conformal metric velocity \cite{smarryork}. This is exactly what happens in spatially homogeneous cosmology where the lapse is spatially constant \cite{bobcmp,bobaihp}. 

However, it may be appropriate \cite{jimmy} to alter the definition of the minimal distortion shift by including the lapse in the spatial volume integration in the variational principle in parallel with the new approach to the vector potential equation \cite{py03} (see their remark after their Eq.~(4.6)), as well as using the proper time conformal velocity. This change in the variational principle from
\beq
  \int \dot {\bar g}{}^{\rm(TF)}{}^a{}_b   \dot {\bar g}{}^{\rm(TF)}{}^b{}_a 
    \,\bar g^{1/2}\omega^1\wedge\omega^2\wedge\omega^3 
\eeq 
to
\begin{eqnarray}
 && \int \overcirc {\bar g}{}^{\rm(TF)}{}^a{}_b   \overcirc {\bar g}{}^{\rm(TF)}{}^b{}_a 
    \,\bar N \bar g^{1/2}\omega^1\wedge\omega^2\wedge\omega^3 
\\
&&=   \int (-2 \bar A{}^a{}_b + \bar N{}^{-1} [\bar L\beta]^a{}_b)   
         (-2 \bar A^b{}_a + \bar N{}^{-1}  [\bar L\beta]^a{}_b)
    \,\bar N \bar g^{1/2}\omega^1\wedge\omega^2\wedge\omega^3 
\nonumber
\end{eqnarray} 
leads to a net factor of $N^{-1}$ in the new variational integral, so that upon variation with respect to $\beta^a$ and integrating by parts (provided the boundary terms can be ignored in a general spacetime), one has this extra factor of $\bar N{}^{-1}$ in the new minimal distortion shift equation, making it identical to the new vector potential equation for generic spacetimes, which is very satisfying, revealing a nice compatibility between the initial value problem decomposition and the dynamical evolution gauge-fixing that would be valid in general spacetimes. In the new minimal distortion gauge, the magnitude of the proper time derivative of the conformal metric is minimized over a thin sandwich between two nearby time coordinate hypersurfaces with the spacetime volume element as the measure over this 4-dimensional region.

\section{The Spatially Homogeneous Case}

The spatially homogeneous spacetimes \cite{ryanshepley} have served a valuable role as a testing ground for general relativity. Restricting the variational approach to Einstein's equations to the minisuperspace of such spacetimes of a fixed Bianchi symmetry type \cite{bianchi} requires examining the boundary terms for troubles \cite{mactaub}. The same boundary term considerations can lead to a disconnect between the supermomenta as constraint equations and as generators of (symmetry compatible) spatial diffeomorphisms in this context \cite{bobvpc80}. In order not to be distracted by the many cases which can arise, it suffices to consider the ``trouble-free" Bianchi type VIII and IX spacetimes where the symmetry specialization does not introduce any such problems since spatial divergences of spatially homogeneous vector fields automatically vanish, making all troublesome spatial boundary terms zero, while remaining general enough within the spatially homogeneous spacetimes to not suffer from other problems \cite{bobaihp}.

The conformal transformation is naturally identified with the usual decomposition of the metric into its unimodular piece and over all determinant. Using matrix notation
\beq
   \bfg = e^{2\beta^0} \tilde{\bfg} \ ,\ 
\det\tilde{\bfg} =1\ ,\
 g = e^{6\beta^0}\ .
\eeq
Without introducing the barred/unbarred notation to complicate matters, we can associate the conformal factor with
\beq
\phi^4=e^{2\beta_0}\ ,\
\phi=e^{\beta_0/2}\ ,\
\phi^{-6}=e^{-3\beta_0}
\ .
\eeq

The unimodular part (conformal metric) is in turn decomposed into a diagonal metric matrix and a linear transformation associated with the action of the natural (symmetry compatible) restricted diffeomorphism group on the spatial metric expressed in a computational frame composed of spatially homogeneous frame vectors \cite{bobcmp}
\beq
  \tilde \bfg = \bfS^T \tilde{\bfg}_{\rm(D)} \bfS\ ,\
  \tilde \bfg_{\rm(D)} =e^{2\breve{\bfbeta}}\ ,\
   \det\bfS = 1\ ,\
  \rm{Tr}\, \breve{\bfbeta} =0\ ,\
   \breve{\bfbeta} = \beta^\pm \bfe_\pm\ ,
\eeq
where
$\bfg_{\rm(D)} = e^{2\bfbeta}$ and $\bfbeta=\beta^0 \bfone +\beta^\pm \bfe_\pm$ (sum implied over $+$, $-$)
decomposes the logarithm of the the full diagonal metric matrix into its pure trace and tracefree parts using the Misner decomposition \cite{misner70,misner69}.

If we identify the space sections of the spatially homogeneous spacetime with the Lie symmetry group (simply transitive case), then
the spatial diffeomorphism freedom compatible with the use of invariant computational frames is the associated group of automorphisms and translations of that group, which induces the action of the linear automorphism group on the space of spatially homogeneous tensor fields. For the semisimple case of Bianchi types VIII and IX, there are only inner automorphisms, also called adjoint transformations, so the action of the linear adjoint group on spatially homogeneous tensor fields is relevant. This group is 3-dimensional (isomorphic to $SO(2,1)$ and $SO(3)$ respectively) and when the spatial structure constants of the invariant frame are chosen to have diagonal form, it acts on the space of spatial metric matrices tranversally to the diagonal metric matrices and so can be used to decompose the metric matrix. $S$ belongs to this group, which leaves invariant the spatial structure constants
\beq
    S^c_l C^l{}_{mn} S^{-1}{}^m{}_a  S^{-1}{}^n{}_b = C^c{}_{ab}\ .
\eeq
The basis of the adjoint Lie algebra is
\beq
[\bfk_a]^b{}_c = C^b{}_{ac}\ ,\
[\bfk_a,\bfk_b] = C^c{}_{ab} \bfk_c\ ,\
\bfS k_a \bfS^{-1} = k_b S^b{}_a\ .
\eeq

The metric parametrization $\bfg = S^T \bfg_{\rm(D)} S$ can be thought of as a transformation from the initial gauge (usually taken to be zero-shift gauge) to diagonal gauge \cite{bobcmp}, where the spatial metric is diagonal and the new spatial frame is $e_{a'} = e_b S^{-1}{}^b{}_a$. 
These new frame vectors must commute with the nonspatial computational frame vector $e_{0'} = \partial/\partial t - B^{a'} e_{a'}$ in order to define a computational frame, which leads to the condition
\beq
  \dot\bfS \bfS^{-1} = \dot{\tilde{\omega}}{}^a \bfk_a = B^{a'} \bfk_a
\rightarrow B^{a'} = \dot{\tilde{\omega}}{}^a
\eeq
for the new shift vector field $B^{a'} e_{a'}$, identifying its new components with the ``autmorphism velocities." These enter the extrinsic curvature or metric velocities in the following way
\beq
 -\frac1{2N}  \bfg^{-1} \dot\bfg 
= \frac1{N} \bfS^{-1} [-\dot\beta^0\bfone -\dot{\breve{\bfbeta}} 
          + \dot{\tilde{\omega}}{}^a \mathcal{K}_a ] \bfS\ ,
\eeq
where the matrix
\beq
\mathcal{K}_a = \frac12(k_a + e^{-2\breve\bfbeta} k_a^T e^{2\breve\bfbeta})\ , 
\eeq
is just the symmetric part of the adjoint matrix $\bfk_a$ with respect to the diagonal spatial metric. The second term in the metric velocity, containing the factor $\dot{\breve{\bfbeta}}$, is the transverse part of the conformal metric velocity, while the third term is its longitudinal part. 

If one starts with zero shift gauge, then the extrinsic curvature matrix is $\bfK =-(2N)^{-1} \bfg^{-1} \dot\bfg$,
and one can interpret this choice of metric parametrization as pointing towards diagonal gauge (for which the conformal metric velocity is transverse), with the vector potential for the tracefree extrinsic curvature defining the associated shift vector field for that new gauge. This was noted in the ``old" conformal approach decomposition where the factor of $\sigma$ was omitted \cite{bobaihp} leading to the outdated identification 
$\beta^a =2\bar N \phi^{-6} Y^a$ instead of the much nicer result $\beta^a=Y^a$ in the new improved approach.
Thus the new approach to the supermomentum constraint and minimal distortion shift gauge finds complete resonance with the mathematical structure of spatially homogeneous dynamics, and suggests extending it to general spacetimes by redefining the minimal distortion shift condition.

\section{Conclusions and Acknowledgments}

Spatially homogeneous spacetimes have provided good feedback to the general theory in many ways over the past half century since they were born at the hands of G\"odel \cite{godel} and Taub \cite{taub}. It was really the general Smarr-York lapse and shift gauge analysis \cite{smarryork} which crystallized my own work in this field as a graduate student under Abe Taub \cite{bobcmp} just before his retirement and led to my first postdoctoral position with Jimmy York in Chapel Hill in 1978. This final nice piece of mathematically pleasing special circumstances for the conformal game that occurs for these spacetimes as I have just described perhaps could have contributed to arriving at the current state of affairs in general a long time ago if I had thought to go further in that direction than I did \cite{bobaihp} and interact more with Jimmy during my year with him. I had not yet learned the lessons of my long association with relativity in Italy that followed beginning the next year with my second postdoctoral position in Rome with Remo Ruffini. We are not in this game just because of the mathematical or physical interest it has for us, but also for the human relationships that it allows us to foster, especially those crossing national boundaries, a career perk that we have been very fortunate to have. 

Giorgio Ferrarese has helped keep alive this spirit of social ties among colleagues that is typically Italian, especially with his tradition of Elba conferences. I was fortunate to be able to accompany Abe Taub on his last visit to Italy at the 1989 such conference  (in menory of Carlo Cattaneo) where I recall a pleasant conversation with Yvonne on the beach. I am happy to have this opportunity again to see her in person and convey my appreciation for her long and fruitful career that has continued to cross paths with Jimmy's and which recently won them jointly the 2003 Heineman Prize for Mathematical Physics (announced in 2002) awarded by the American Physical Society and the American Institute of Physics. In a way this helped bring Jimmy and me together again in Rio last summer by catalyzing his attendance at the Tenth Marcel Grossmann Meeting (2003) through my serendipitous intervention and subsequently reawakened my interest in his work that produced this investigation last fall; I thank Jimmy for his helpful feedback on this manuscript.
Finally I especially thank Giorgio for his warm hospitality over the years that brought us all together in Porto Azzuro, Elba in June of 2004.


\end{document}